\documentclass[twocolumn, amsmath,amssymb,aps]{revtex4-1}

\usepackage{verbatim}
\usepackage{graphicx}
\usepackage{bm} 
\usepackage{ifpdf} 
\usepackage{dcolumn}  
\usepackage{epsfig}
\usepackage{color}
\usepackage{amsmath}
\usepackage[usenames,dvipsnames]{xcolor}

\makeatletter 
\renewcommand\@biblabel[1]{#1.} 
\makeatother

\begin{document}

\title{Derivation of the Inverse Schulze-Hardy Rule}
\author{Gregor Trefalt}
\thanks{E-mail: \texttt{gregor.trefalt@unige.ch}}
\affiliation{Department of Inorganic and Analytical Chemistry, University of Geneva, Sciences II, 30 Quai Ernest-Ansermet, 1205 Geneva,
Switzerland}

\date{\today}

\begin{abstract}
The inverse Schulze-Hardy rule was recently proposed based on experimental observations. This rule describes an interesting situation of the aggregation of charged colloidal particles in the presence of the multivalent {\sl coions}. Specifically, it can be shown that the critical coagulation concentration is inversely proportional to the coion valence. Here the derivation of the inverse Schulze-Hardy rule based on purely theoretical grounds is presented. This derivation complements the classical Schulze-Hardy rule which describes the multivalent {\sl counterion} systems.
\end{abstract}

\maketitle

The aggregation of charged colloids is a long-studied phenomenon. More than 100 years ago Schulze and Hardy showed that the aggregation power of salts depends strongly on the ion valence~\cite{Schulze1882, Hardy1900}. More precisely, the critical coagulation concentration (CCC) (i.e., concentration of salt at which particles start to aggregate fast)~\footnote{Colloids are typically stable at low salt concentrations and unstable at higher ones. The sharp transition between these two regimes is defined as the CCC. At this concentration the energy barrier, which prevents the aggregation at lower concentration, vanishes.} decreases very rapidly by increasing {\sl counterion} valence. This discovery was later confirmed theoretically by Derjaguin, Landau, Verwey and Overbeek and is known as the DLVO theory~\cite{Derjaguin1941, Verwey1948}. They have shown that by assuming the interaction between particles as a sum of van der Walls (vdW) and double layer forces (DL) in the symmetric $z$:$z$ electrolyte the CCC is inversely proportional to the sixth power of the valence,
\begin{equation}
{\rm CCC} \propto \frac{1}{z^6} \,\,\,\, \text{(Schulze-Hardy rule).}
\label{eqn:SH}
\end{equation}
The above relation, also named Schulze-Hardy rule, is valid for highly charged particles. This explanation confirmed the DLVO theory and made it widely accepted.

The symmetric $z$:$z$ electrolytes are usually practically insoluble, therefore in experiments one typically uses asymmetric $1$:$z$ or $z$:$1$ multivalent electrolytes. In the case of asymmetric electrolytes, multivalent ions can either play a role of the {\sl counterions} or the {\sl coions}, where they have the opposite or the same charge as the colloidal particle, respectively. It was shown experimentally as well as theoretically that for highly charged particles the Schulze-Hardy rule (\ref{eqn:SH}) is a good approximation also for asymmetric electrolytes where $z$ is the counterion valence~\cite{Oncsik2014,Hsu1995,Trefalt2013}. Recently Cao {\sl et al.}~\cite{Cao2015} investigated a complementary problem, namely the influence of multivalent {\sl coions} on the aggregation. In this situation experimental data could be reasoned with the {\sl inverse Schulze-Hardy rule}, namely
\begin{equation}
{\rm CCC} \propto \frac{1}{z} \,\,\,\, \text{(inverse Schulze-Hardy rule),}
\label{eqn:ISH}
\end{equation}
where $z$ is the valence of the coion. Note that in the case of the coions the dependence on valence is much weaker. Interestingly, in the low particle charge limit, where the DL forces can be described by the Debye-H\"uckel (DH) approximation, the same
\begin{equation}
{\rm CCC} \propto\frac{1}{z(z+1)}
\label{eqn:DH}
\end{equation}
dependence for both counterions and coions is reached~\cite{Trefalt2013, Cao2015}. The latter low charge limit (\ref{eqn:DH}) lies between the Schulze-Hardy (\ref{eqn:SH}) and inverse Schulze-Hardy (\ref{eqn:ISH}) dependences. This proposed inverse Schulze-Hardy rule therefore elegantly completes the understanding the aggregation in experimentally relevant asymmetric multivalent electrolytes.

The Schulze-Hardy rule (\ref{eqn:SH}) and the low charge DH limit (\ref{eqn:DH}) were both derived theoretically. On the other hand, the inverse Schulze-Hardy rule was only given as an empirical dependence based on experimental observations.~\cite{Cao2015} Therefore, the aim of this paper is to present the  derivation of the inverse Schulze-Hardy rule based solely on theoretical grounds.

A naive explanation of the inverse Schulze-Hardy rule would come from the original argument of Schulze and Hardy. They have explained that the CCC is controlled by the counterion concentration. In asymmetric $1$:$z$ and $z$:$1$ electrolytes, where $z$ represents the coions, the counterions are monovalent. In these systems, the concentration of monovalent counterions is equal to $zc$, where $c$ is the salt concentration. If one now assumes that the aggregation happens at constant counterion concentration, this leads to the $1/z$ dependence of the salt concentration. In this situation, ${\rm CCC}\propto 1/z$ proportionality stems solely from the composition of the $1$:$z$ and $z$:$1$ salts. However, as it will be shown below, this simple intuitive reasoning cannot be justified.

\begin{figure}[t]
\small
\centering
\includegraphics[width=8.5cm]{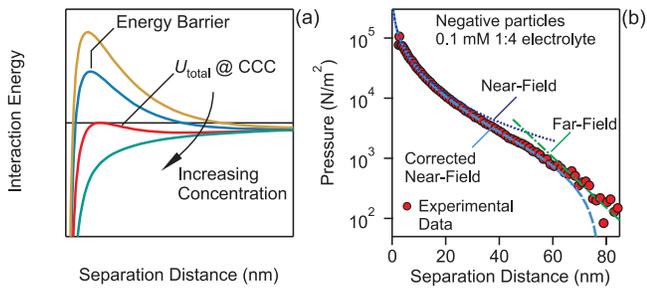}
\caption{(a) Schematic presentation of interaction energy evolution with increasing salt concentration. (b) Disjoining pressure between two negatively charged particles in the presence of 0.1 mM 1:4 electrolyte, experimental data taken from~\cite{MontesRuiz-Cabello2015}. Full Poisson-Boltzmann, near-field, corrected near-field, and far-field curves are also presented. Note that in the presented case the van der Waals interactions are negligible.}
\label{fig:forcesCoions}
\end{figure}

\begin{figure}[t]
\small
\centering
\includegraphics[width=8.5cm]{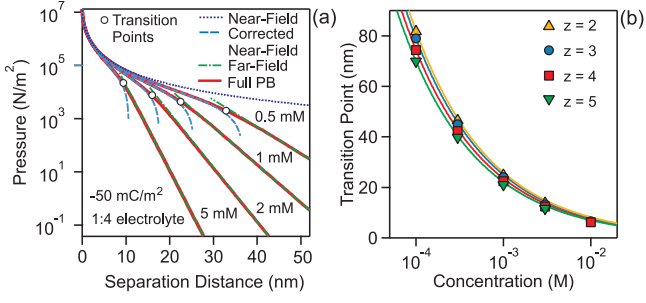}
\caption{(a) Evolution of disjoijning pressures between two negatively charged plates with increasing 1:4 salt concentration. Full PB solution is presented by full lines, corrected near-field limit dashed lines, and far-field limit dashed-dotted lines. (b) Transition point as a function of salt concentration for multivalent coions of valence $z$. Points present the full PB calculations, lines are the approximation from the thermal condition~(\ref{eq:thermal3}). Note that only the electrostatic part of the interactions is presented. Surface charge density of $-50$~mC/m$^2$ is used.}
\label{fig:forcesNearFarField}
\end{figure}

In order to describe the aggregation in the presence of multivalent coions one can follow the original DLVO approach. This derivation is based on the calculation of the total interaction energy $U_{\rm total}$, between two charged colloids as a sum of attractive vdW and repulsive DL contributions,
\begin{equation}
U_{\rm total} = U_{\rm vdW} + U_{\rm DL}.
\label{eq:DLVO}
\end{equation}
At low salt concentrations the DL interactions are dominant and an energetic barrier develops. With increasing concentration the barrier diminishes and when it is close to zero the particles aggregate, see Fig.~\ref{fig:forcesCoions}a. The salt concentration at which the energy barrier vanishes is a good approximation for the CCC. One can therefore understand the effect of multivalent ions on aggregation through the effect of such ions on the interactions. The interactions in the presence of multivalent counterions were studied extensively by both experimentalists~\cite{Pashley1984, Besteman2004, Zohar2006, Dishon2011, Sinha2013, MontesRuiz-Cabello2014a, MontesRuiz-Cabello2014b, MoazzamiGudarzi2015} and theoreticians~\cite{Kjellander1990, Wu1999, Moreira2001, Trulsson2006, Kanduc2011, Hatlo2009, Bohinc2012}. In particular the effect of ion correlations and validity of the mean-field Poisson-Boltzmann (PB) treatment was addressed in these studies. On the other hand, the interactions between charged particles in the presence of multivalent coions, which are of interest here, received much less attention~\cite{Kohonen2000, MontesRuiz-Cabello2015, Wu1999}. In these situations the double-layer interactions are much softer and longer-ranged as compared to the interactions in the presence of monovalent electrolytes or multivalent counterions. The multivalent coions also have a profound influence on the shape of the force-curves. While in the presence of monovalent electrolyte and multivalent counterions the profiles are exponential down to small separations, the shape of the curves for multivalent coions is exponential only at large separations. In the latter case the interaction can be decomposed into near-field algebraic and far-field exponential parts~\cite{MontesRuiz-Cabello2015}. Fig.~\ref{fig:forcesCoions}b shows such interaction between two negatively charged particles in the presence of 1:4 electrolyte.
Note that both the particle and the multivalent ion are negatively charged. An oppositely charged system with positively charged particles and multivalent cations can be also realized~\cite{MontesRuiz-Cabello2015}. The experimental profile can be accurately described by mean-field PB theory. In this case, no ion-correlations effects are expected~\cite{Wu1999}. The PB equation is solved numerically for two charged plates immersed in an asymmetric electrolyte solution, for details see~\cite{Trefalt2013, MontesRuiz-Cabello2015}. The characteristic shape of the pressure curve is the consequence of expulsion of multivalent coions from the gap between the surfaces, which in the presented case happens at about 60 nm, see Fig.~\ref{fig:forcesCoions}b. The near-field pressure $\Pi_{\rm DL}^{\rm near}$, therefore corresponds to the monovalent counterion only case (i.e., multivalent coions are expelled) and can be approximated as~\cite{Israelachvili2011, Briscoe2002a, MontesRuiz-Cabello2015}
\begin{equation}
\Pi_{\rm DL}^{\rm near} (h) = \frac{2\pi^2\varepsilon\varepsilon_0}{\beta^2e_0^2}\cdot\frac{1}{h^2},
\label{eq:nearField}
\end{equation}
where $\varepsilon\varepsilon_0$ is the dielectric permittivity, $\beta=1/kT$ is the inverse thermal energy, $e_0$ is the elementary charge, and $h$ is the separation distance. The above expression is valid for the distances larger than the Gouy-Champman length $\lambda = \frac{2\varepsilon\varepsilon_0}{\beta e_0 \sigma}$ with $\sigma$ being the surface charge density. The counterion only case sets the concentration of salt outside the gap to zero. However, in the case of the added 1:$z$ salt, as in Fig.~\ref{fig:forcesCoions}b, the concentration outside the gap is finite. The near-field pressure can therefore be corrected for the osmotic pressure outside the gap
\begin{equation}
\Pi_{\rm DL}^{\rm near} (h) = \frac{2\pi^2\varepsilon\varepsilon_0}{\beta^2e_0^2}\cdot\frac{1}{h^2}-\frac{(z+1)c}{\beta},
\label{eq:nearFieldCorrected}
\end{equation}
where $c$ is the bulk concentration of 1:$z$ salt. Note that this simple correction substantially improves the accuracy of the near-field approximation at larger distances, see Fig~\ref{fig:forcesCoions}b. The far-field pressure $\Pi_{\rm DL}^{\rm far}$, is of DH type~\cite{Israelachvili2011, MontesRuiz-Cabello2015}
\begin{equation}
\Pi_{\rm DL}^{\rm far} (h) = 2\varepsilon\varepsilon_0\kappa^2\psi_{\rm eff}^2 e^{-\kappa h},
\label{eq:farField}
\end{equation}
where $\psi_{\rm eff}$ is the effective surface potential, $\kappa=\sqrt{\frac{2\beta e_0^2 I}{\varepsilon\varepsilon_0}}$ is the inverse Debye length, and $I$ is the ionic strength calculated as $I=\frac{1}{2}z(z+1)c$ for 1:$z$ electrolyte. Note that the effective potential for asymmetric electrolytes is not known analytically. We can now define a transition point $h_{\rm t}$ between the the two limits. Below and above the $h_{\rm t}$, near-field and far-field limits are valid, respectively. Such treatment successfully describes the experimental force-curves~\cite{MontesRuiz-Cabello2015}. By increasing the concentration of salt the near-field limit is unaffected, what changes are the far-field limit and the transition region with the transition point, see Fig~\ref{fig:forcesNearFarField}a.
The fact that the near-field~(\ref{eq:nearField}) does not depend on monovalent counterion concentration rules out the simple intuitive explanation of the inverse Schulze-Hardy rule which is described above. The CCC cannot be controlled solely by counterion concentration as the near-field interaction does not depend on it in the case of multivalent coions. To get the complete picture one has to rather look at how the transition point and far-field behavior are affected by the addition of salt.

The transition point $h_{\rm t}$, between near-field and far-field approximations represents a separation distance, which marks the start of expulsion of the multivalent co-ions from the region between the two charged plates upon closer approach. The exclusion happens due to electrostatic repulsion between charged plates and the coion. The interaction between a coion and the two plates can be estimated by calculating the electrostatic potential at the mid-plane between two plates. When the plates are closer than $h_{\rm t}$ only counterions are present in the slit. Here we assume positive plates with monovalent anions as counterions. Note that the analogous situation with negative plates and monovalent cations is possible and would yield the same result. For the former case of positively charged plates the following form of the PB equation has to be satisfied~\cite{Israelachvili2011, MontesRuiz-Cabello2015, Shklovskii1999}
\begin{equation}
\frac{{\rm d}^2 \psi}{{\rm d} x^2}  = \frac{e_0c_-}{\varepsilon\varepsilon_0}e^{\beta e_0 \psi},
\label{eq:PoissonBoltzmann}
\end{equation}
where $\psi$ is the electric potential and $c_-$ is the number concentration of the counterions. The prefactor $c_-$ fixes the potential at the surface of the plate to the value of surface potential in $z$:1 electrolyte in the limit of high surface charge densities.The solution of Eq.~(\ref{eq:PoissonBoltzmann}) gives the electric potential at the mid-plane
\begin{equation}
\psi_M (h) = -\frac{2}{\beta e_0}\ln \left( \frac{\alpha h}{2\pi} \right),
\label{eq:midPlanePotential}
\end{equation}
Now the multivalent coion with valence $z$ wants to enter the slit and is affected by the electrostatic repulsion exerted by the plates. One can approximate that the coion will enter the region between the two plates when the electrostatic energy at the mid-plane is equal to 2~$kT$
\begin{equation}
z\beta e_0 \psi_{\rm M}= 2.
\label{eq:thermal}
\end{equation}
At this point the separation of the two plates is equal to $h_{\rm t}$. Combining Eq.~(\ref{eq:midPlanePotential}) and Eq.~(\ref{eq:thermal}) yields the position of the transition point
\begin{equation}
h_{\rm t} = \frac{2 \pi}{\alpha}e^{-\frac{1}{z}} .
\label{eq:transitionPoint1}
\end{equation}
In the case of asymmetric $z$:1 electrolyte, where $c_- = zc$ and $I = \frac{1}{2}z(z+1)c$ we finally arrive at
\begin{equation}
 h_{\rm t} \kappa = 2\pi\sqrt{\frac{z+1}{2}}e^{-\frac{1}{z}} .
\label{eq:thermal3}
\end{equation}
The variation of the transition point with $z$:1 electrolyte concentration is shown in Fig.~\ref{fig:forcesNearFarField}b. One can observe that the results from the full PB treatment can be well approximated with relation (\ref{eq:thermal3}).

By knowing the position of the transition point, we can approximate the electrostatic interaction between the particles by using the near-field limit (\ref{eq:nearField}) when $h\le h_{\rm t}$ and the far-field limit (\ref{eq:farField}) for $h>h_{\rm t}$. The interaction force between two particles with radius $R$, can then be obtained by integration of the pressure and by application of the Derjaguin approximation
\begin{equation}
F_{\rm DL}^{\rm near}=\pi R \int_{h}^{h_{\rm t}} \Pi_{\rm DL}^{\rm near}(h'){\rm d}h' + \pi R \int_{h_{\rm t}}^\infty \Pi_{\rm DL}^{\rm far}(h'){\rm d}h' ,
\label{eq:forceNear1}
\end{equation}
which yields near-field force
\begin{multline}
\frac{F_{\rm DL}^{\rm near}}{\pi R}= \frac{2\pi^2\varepsilon\varepsilon_0}{\beta^2 e_0^2} \left( \frac{1}{h} - \frac{1}{h_{\rm t}} \right) + \frac{\varepsilon\varepsilon_0 \kappa^2}{\beta^2 e_0^2 z}\left( h- h_{\rm t} \right) \\
+ 2\varepsilon\varepsilon_0 \kappa \psi_{\rm eff}^2 e^{-\kappa h_{\rm t}} .
\label{eq:forceNear2}
\end{multline}
By analogy the far-field force is
\begin{equation}
\frac{F_{\rm DL}^{\rm far}}{\pi R} = 2\varepsilon\varepsilon_0 \kappa \psi_{\rm eff}^2 e^{-\kappa h}.
\label{eq:forceFar1}
\end{equation}
Integration of the force yields the potential energy profile in the near-field limit
\begin{equation}
U_{\rm DL}^{\rm near} = \int_{h}^{h_{\rm t}} F_{\rm DL}^{\rm near}(h'){\rm d}h' + \int_{h_{\rm t}}^{\infty} F_{\rm DL}^{\rm far}(h'){\rm d}h' .
\label{eq:energyNear1}
\end{equation}
The two integrals above can be solved analytically yielding the expression
\begin{multline}
\frac{U_{\rm DL}^{\rm near}}{\pi R} = B\ln \left( \frac{h_{\rm t}}{h} \right)\\
+ B\left(1-\frac{h}{h_{\rm t}}\right) \left( \frac{1}{\kappa h_{\rm t}} - 1 - \frac{\kappa h_{\rm t}}{2\pi^2 z}-\frac{\kappa^2 h_{\rm t}^2}{2\pi^2 z}\right)\\
+ B\left(1-\frac{h^2}{h_{\rm t}}\right)\left( \frac{\kappa^2 h_{\rm t}^2}{4\pi^2 z} \right) + \frac{B}{\kappa^2h_{\rm t}^2} - \frac{B}{2\pi^2 z}, 
\label{eq:energyNear2}
\end{multline}
where constant $B = \frac{2\pi^2\varepsilon\varepsilon_0}{\beta^2 e_0^2}$ and the equality $\Pi_{\rm DL}^{\rm near} (h_{\rm t}) = \Pi_{\rm DL}^{\rm far} (h_{\rm t})$ were used to write the equation in the condensed form.

\begin{figure}[t]
\small
\centering
\includegraphics[width=8.5cm]{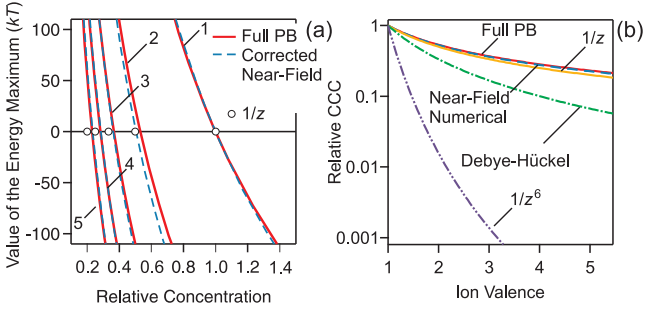}
\caption{(a) Value of the energy maximum in the interaction total energy potential between two charged particles as a function of relative concentrations of multivalent coions. Results for the valence of the coions between 1 and 5 are presented. The concentrations are normalized by concentration of 1:1 electrolyte where the energy barrier is 0 $kT$. Curves for the full PB solution and the near-field solution (\ref{eq:energyNear2}) are presented. (b) Relative CCC as a function of coion valence calculated with the full PB, the near field solution and the analytically derived inverse Schulze-Hardy $1/z$ dependence. Low-potential Debye-H\"uckel and Schulze-Hardy $1/z^6$ dependences are also shown. Surface charge density of 0.4 C/m$^2$ was used for PB calculations. Hamaker constant of 1$\cdot$10$^{-18}$~J and $R = 150$~nm were used throughout.}
\label{fig:coionCCC}
\end{figure}

The total interaction energy (\ref{eq:DLVO}) can be now calculated by summing the vdW and DL energies. The van der Waals contributions can be approximated by simple non-retarded expresions
\begin{gather}
F_{\rm vdW}= -\frac{HR}{12}\cdot\frac{1}{h^2}, \\
U_{\rm vdW} = -\frac{HR}{12}\cdot\frac{1}{h},
\label{eq:vdW}
\end{gather}
where $H$ is the Hamaker constant. Following the original DLVO condition~\cite{Derjaguin1941, Verwey1948} the CCC can be estimated by setting the energy barrier of the total interaction energy to zero. This condition can be mathematically written as
\begin{equation}
\left . \frac{dU_{\rm total}}{dh}  \right |_{h_{\rm max}}= - F_{\rm total} (h_{\rm max}) = 0 \,\,\,\, {\rm and} \,\,\,\, U_{\rm total} (h_{\rm max}) = 0 
\label{eq:SHConditon}
\end{equation}
The full PB solution for the interaction energies between two charged particles in the presence of multivalent coions shows that the position of the maximum in the energy profile is generally at smaller separations as compared to the position of the transition point, $h_{\rm max} < h_{\rm t}$. Therefore, Eqs.~(\ref{eq:thermal3}), (\ref{eq:forceNear2}--\ref{eq:forceFar1}), (\ref{eq:energyNear2}--\ref{eq:vdW}) can be used with condition (\ref{eq:SHConditon}) to numerically calculate the evolution of the value of the energy maximum with the concentration within the near-field approximation, see Fig.~\ref{fig:coionCCC}a.
At the concentration when the energy maximum reaches zero (i.e., the barrier is 0 $kT$) the particles aggregate. The numerical near-field solution confirms that $h_{\rm t} \gg h_{\rm max}$ and that $h_{\rm max}$ at CCC is practically independent of the coion valence. We can use these criteria to further approximate the near-field energy limit and calculate the total energy at maximum using Eq.~(\ref{eq:energyNear2}) and Eq.~(\ref{eq:vdW})
\begin{equation}
\frac{U_{\rm total}(h_{\rm max})}{\pi R} = B\ln \left( \frac{h_{\rm t}}{h_{\rm max}} \right) +Bf(z) - \frac{H}{12\pi h_{\rm max}},
\label{eq:energyNearApprox}
\end{equation}
where $f(z)$ is a function of coion valence
\begin{equation}
f(z) = \left( \frac{1}{\kappa h_{\rm t}} - 1 - \frac{\kappa h_{\rm t}}{2\pi^2 z}-\frac{\kappa^2 h_{\rm t}^2}{4\pi^2 z} + \frac{1}{\kappa^2 h_{\rm t}^2} - \frac{1}{2\pi^2z} \right)\, ,
\label{eq:valenceFunction}
\end{equation}
which can be well approximated with $f(z) \approx \frac{3}{2} -\frac{1}{z}$.
Let us now calculate the concentration at which the energy barrier vanishes, this concentration corresponds to the CCC. First we can get the following expression for $h_{\rm t}$ by equating Eq.~(\ref{eq:energyNearApprox}) to zero
\begin{equation}
h_{\rm t} = Ce^{-\frac{1}{z}},
\label{eq:transitionPointNear}
\end{equation}
where $C = h_{\rm max} \exp\left(\frac{H}{12\pi B h_{\rm max}}+\frac{3}{2} \right)$ is a constant. By using Eq.~(\ref{eq:thermal3}) we arrive at
\begin{equation}
\kappa = \frac{2\pi\sqrt{\frac{z+1}{2}}e^{-\frac{1}{z}}}{Ce^{-\frac{1}{z}}}\, .
\label{eq:kappa1}
\end{equation}
For a 1:1 electrolyte ($z=1$) Eq.~(\ref{eq:kappa1}) yields
\begin{equation}
\kappa \approx \frac{2\pi}{C}
\label{eq:kappa11}
\end{equation}
and for a $z$:1 electrolyte ($z>1$) with
\begin{equation}
\kappa \approx \frac{2\pi}{C}\sqrt{\frac{z+1}{2}}
\label{eq:kappaz1}
\end{equation}
At the CCC the Debye length for $z$:1 electrolyte is defined as
\begin{equation}
\kappa^2 = \frac{2\beta e_0^2}{\varepsilon\varepsilon_0}\cdot\frac{z(z+1)}{2}\cdot{\rm CCC}
\label{eq:kappa1z}
\end{equation}
Combining Eqns.~(\ref{eq:kappa11})--(\ref{eq:kappa1z}) we arrive at the expressions for CCC in 1:1 electrolyte
\begin{equation}
{\rm CCC} (z=1) = \frac{2\pi^2\varepsilon\varepsilon_0}{C^2\beta e_0^2},
\label{eq:CCC11}
\end{equation}
and in $z$:1 electrolyte for $z>1$
\begin{equation}
{\rm CCC} (z>1) = \frac{2\pi^2\varepsilon\varepsilon_0}{C^2\beta e_0^2}\cdot \frac{1}{z}.
\label{eq:CCCz1}
\end{equation}
From Eq.~(\ref{eq:CCC11}) and Eq.~(\ref{eq:CCCz1}) the inverse Schulze-Hardy rule immediately follows
\begin{equation}
\frac{\rm CCC}{{\rm CCC} (z=1)} = \frac{1}{z}.
\label{eq:inverseSHFinal}
\end{equation}
Finally, in Fig.~\ref{fig:coionCCC}b the relative CCC as a function of ion valence is plotted. One can observe that the full PB and near-field numerical solutions match closely the inverse Schulze-Hardy $1/z$ dependence. These curves represent the high-charge limit and their dependence is weaker as compared to the low-charge Debye-H\"uckel limit. The case of multivalent counterions, which yields much stronger Schulze-Hardy $1/z^6$ dependence in the high-charge limit, is also shown.

In conclusion, the derivation of the inverse Schulze-Hardy rule is shown. The simple inverse proportionality of the CCC on the coion valence is not caused by monovalent ion concentration as one would naively expect but rather by the interplay between counterion only and Debye-H\"uckel type of interactions. The transition point between near-field and far-field regime, at which coions begin to be expelled from the slit, turns out to be critical for understanding the aggregation process. This work complements the classical Schulze-Hardy rule and extends our understanding of aggregation in multivalent asymmetric electrolytes. Now both, counterion and coion, high-charge limits are explained. Furthermore, the low-charge Debye-H\"uckel limit is the same for counterions and coions, as the weakly charged surface does not distinguish between counterions and coions. I hope that these results will stimulate further research on the use of multivalent coions in tuning the stability of colloids and could be possibly used for colloidal self-assembly.

I acknowledge fruitful discussions with Michal Borkovec, Svilen Kozhuharov, and Du\v sko \v Cakara and financial support from the University of Geneva and the Swiss National Science Foundation.

\bibliography{paperLib}

\bibliographystyle{rsc}

\end{document}